# Noise Figure Evaluation Using Low Cost BIST

Marcelo Negreiros, Luigi Carro, Altamiro A. Susin
*Instituto de Informática - Programa de Pós Graduação em Computação - PPGC*
*Universidade Federal do Rio Grande do Sul - UFRGS, Porto Alegre, RS, Brazil*
negreiro,carro,susin@inf.ufrgs.br

## Abstract

*A technique for evaluating noise figure suitable for BIST implementation is described. It is based on a low cost single-bit digitizer, which allows the simultaneous evaluation of noise figure in several test points of the analog circuit. The method is also able to benefit from SoC resources, like memory and processing power. Theoretical background and experimental results are presented in order to demonstrate the feasibility of the approach.*

## 1. Introduction

With the increasing capacity of development achieved by the electronic industry nowadays, mainly in the consumer market, the design of new devices is a great challenge. A fast, flexible and reliable methodology must be used in order to cope with ever shrinking product development time, as a shorter time-to-market is crucial to the commercial success of such devices. The SoC approach [1] has been used in order to solve this issue.

Testing of such devices is another challenge, as test costs must be kept lower for the device to be competitive in the market. Unfortunately, analog testers are expensive devices and extra difficulties arise when trying to test analog circuits in a SoC environment. The limited access to the input and output of the analog circuit under test is an example.

In order to reduce the cost of the analog test, built-in test strategies may be used in order to reduce the requirements of external analog testers, or replace them completely. In the SoC environment, as plenty of processing and memory resources are available, it is possible to perform test analysis by reusing these resources [2].

In this paper a low cost method for evaluating the noise figure of analog circuits is presented. Noise figure is an important parameter in the specification and design of low noise systems, such as communications systems and biomedical instrumentation. It is used to characterize the noise behavior, from single analog components to entire systems.

The test method is based on digital signal processing and may be implemented in the SoC environment by reusing processing and memory resources already available in the SoC. A one-bit digitizer that is permanently connected to the desired analog test point is used, thus minimally disturbing the circuit under test. Thanks to the simplicity of the converter, low analog area overhead is obtained, and no impact is made on the noise figure of the circuit being tested, that is, the proposed BIST does not increase the noise level to be measured.

The ultimate goal of this work is to show that, by using a simple BIST cell [3], one can measure not only frequency related parameters of the circuit under test, but rather one can obtain information of other important characteristics like noise figure.

The paper is organized as follows: in section 2 a brief review of other approaches to measure the Noise Figure is presented. In section 3, noise parameters are reviewed and two common noise figure measurement methods are presented. An analysis of a possible implementation of the methods here proposed in the SoC environment is provided in section 4, together with the test method. In section 5 results regarding the digitizer are presented, followed by experimental noise figure evaluations. Analysis is provided in section 6 and the paper finishes with conclusions in section 7.

## 2. Related work

The use of embedded noise sources for noise figure measurements has already been reported. In [4] a thin-film resistor has been used as an on-wafer noise source. Diode noise sources have been investigated in [5]. A production test scheme based on signature testing has been proposed in [7], where noise figure measurements have been indirectly made. To the authors' knowledge, a first approach showing that it is possible to have a BIST circuit to measure noise figure has been proposed in [6]. In the present work we address the issue of using a low cost BIST to measure the noise figure.

## 3. Noise in analog circuits

In this section some background information regarding noise characterization in analog circuits is provided. First noise figure is defined, and then two usual measurement methods are presented.

### 3.1 Definitions

A common parameter used to characterize the noise behavior of an analog electric *signal* (such as the output of a sensor or amplifier) is the signal-to-noise ratio (SNR). It is a ratio of the signal power to the noise power, expressed in dB.

$$SNR = 10 \cdot \log_{10}\left(\frac{V_S^2}{V_N^2}\right) \quad dB \tag{1}$$

Noise figure (NF) and noise factor (F) are parameters used to characterize the noise behavior of a *device or circuit*, as shown in figure 1.

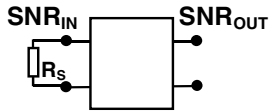

**Figure 1. Noise characterization of a 2-port device**

The noise factor (F) of a two-port device is the ratio of the available output noise power per unit bandwidth to the noise caused by the actual source connected to the input terminals of the device, measured at the standard temperature of 290K [8].

$$F = \frac{SNR_{IN}}{SNR_{OUT}} \tag{2}$$

Noise figure (NF) is defined as the noise factor (F) expressed in dB [8,9]

$$NF = 10 \cdot \log_{10}(F) \quad dB \tag{3}$$

The IEEE standard definition of the noise factor (F) is given by equation 4, where $N_a$ is the noise added by system, $T_0$ is 290K (standard temperature), $B$ is the system bandwidth, $k$ is the Boltzmann constant and $G$ is the gain of the system.

$$F = \frac{N_a + k \cdot T_0 \cdot B \cdot G}{k \cdot T_0 \cdot B \cdot G} \tag{4}$$

The definition of noise figure is based on the assumption of a linear system. Some extensions for non-linear systems have been proposed, like in [10], but are not going to be analyzed in this work.

Some usual noise figure values are illustrated in table 1, together with the corresponding noise factor. Note that the typical value of noise figure for an RF low noise amplifier is 3dB. For an RF mixer, the value is about 10 dB [11]. A circuit that does not add noise to its input would have a noise figure of 0 dB.

**Table 1. Some reference values for noise figure and noise factor**

| NF(dB) | F | Example |
|---|---|---|
| 0 | 1 | noiseless analog circuit |
| 3 | 2 | RF low noise amplifier |
| 10 | 10 | RF mixer |

### 3.2 Measurement methods

In the **direct measurement method**, equation 4 is used. A load is connected to the input of the system, at a temperature of 290K, like in figure 1. The output noise power of the system is measured (the numerator of equation 4). If one knows the measurement bandwidth (**B**) and the gain of the system (**G**), equation 4 can be used directly.

In the **y-factor method**, noise figure evaluation is based on the use of a calibrated noise source [8,12]. The method is a two-step process (see figure 2): with the noise source turned off (at a temperature of 290K, or cold temperature), the DUT output power ($N_c$) is measured. Then the noise generator is turned on, and the noise output power ($N_h$) for the hot temperature is recorded. The Y factor is the ratio of these powers:

$$Y = \frac{N_h}{N_c} \tag{5}$$

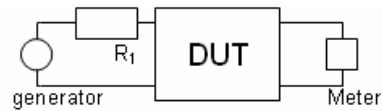

**Figure 2. Noise characterization setup**

Knowing that the noise power at the output of the DUT is simply the noise added by the system ($N_a$) plus the amplified input noise, one can write

$$N_h = k \cdot T_h \cdot B \cdot G + N_a \quad (6)$$

$$N_c = k \cdot T_c \cdot B \cdot G + N_a \quad (7)$$

After applying equations 6 and 7 to 5, and developing using equation 4, one obtains the equation of the Y-Factor technique [12], that allows the evaluation of the noise factor

$$F = \frac{(T_h/T_0 - 1) - Y(T_c/T_0 - 1)}{(Y - 1)} \quad (8)$$

$T_0$ is the reference temperature of 290K. If the noise source cold temperature is not 290K, this provides a correction term. This equation can be rewritten in order to take into account noise powers, instead of temperatures [10], as shown in equation 9.

$$F = \frac{(N_h/N_0 - 1) - Y(N_c/N_0 - 1)}{(Y - 1)} \quad (9)$$

## 4. A NF BIST in the SoC environment

In this section the implementation of methods for estimating noise figure suitable for BIST in a SoC environment are discussed. The proposed method is also presented.

### 4.1 Direct method

An implementation of the direct method would be as shown in figure 3. A nominal load $R_s$ must be applied at the input at a temperature of $T_0$=290K. The output of the DUT must be amplified and routed to the ADC of the system for further processing.

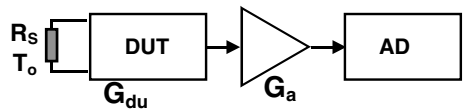

**Figure 3. Direct method setup**

One practical disadvantage of this setup is related to variations in the gain of the amplifier ($G_a$). This gain multiplies both terms in equation 10, but only the numerator is measured. This way, any deviation in the amplifier gain (from $G_a$ to $G_a$') will cause an error in the noise factor estimation. This issue is expected to occur because of process variations that may affect the gain of the amplifier.

$$F = \frac{(N_a + k \cdot T_o \cdot B \cdot G_{DUT})G_a'}{k \cdot T_o \cdot B \cdot G_{DUT} \cdot G_a} \quad (10)$$

### 4.2 Y-factor method

If one could embed a suitable noise generator, being able to provide two known noise levels, it would be feasible to implement a noise figure measuring system based on the Y-factor technique. The system level setup is presented in figure 4. A programmable attenuator provides the noise levels needed for the NF measurement. The generator noise level can be measured through an auxiliary analog path to the ADC.

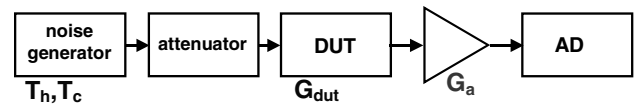

**Figure 4. Y-factor setup**

This setup does not possess the same sensitivity to changes in the amplifier gain as the direct method. Both numerator and denominator in equation 11 are measured, so any deviations in the amplifier gain (from $G_a$ to $G_a$') are corrected.

$$Y = \frac{(N_a + k \cdot T_h \cdot B \cdot G_{DUT})G_a'}{(N_a + k \cdot T_c \cdot B \cdot G_{DUT})G_a'} \quad (11)$$

In [6] an analysis of uncertainty in commercial noise sources is provided, and it is shown that even large errors like 5% in the hot temperature can still provide useful measurements for noise figure estimation, if an error of ±0.3dB is acceptable (for noise figures of 3dB and 10dB). These errors could be used as guidelines in the design of the noise generator and attenuator.

### 4.3 Proposed method for NF measurement

Some problems are not addressed in the strategy presented in figure 4: the AD converter of the system is used, so simultaneous acquisition is not possible. Also, routing of analog signals to the ADC may be difficult. There is a need for a multiplexing device at the input of the ADC, which introduces non-linearity and distortion in the signal.

If the ADC is replaced by a simple digitizer, like in figure 5, some advantages like the possibility of simultaneous observability and no need for multiplexing devices is achieved. Also, because of the simplicity of the digitizer, it can be permanently connected to the analog test point, thus avoiding switches which degrade the performance of the analog circuit under test.

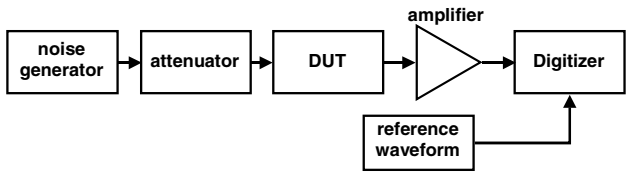

**Figure 5. Proposed setup**

As the Y-factor method requires the evaluation of a ratio of signal power, the digitizer must be able to provide noise power levels.

In this work the digitizer [3] uses a voltage comparator and a reference signal in order to perform the data acquisition. The requirements of the reference signal are modest (in the sense that only a small frequency band is used in the calibration process), allowing a simple and low-cost signal generator to be used.

## 5. Results

In this section the digitizer and a strategy to evaluate power levels using a reference waveform are presented. Noise and reference levels are also discussed. The section finishes with results from a prototyped setup.

### 5.1 Digitizer

The digitizer (figure 6) is composed by a voltage comparator with a noise reference [3]. The input signal is connected directly to the input of the comparator. The digitized input signal is obtained at the output of the comparator and is a digital output. Sampling may be controlled by adding a flip-flop at the output of the comparator. This structure is common in high speed voltage comparators.

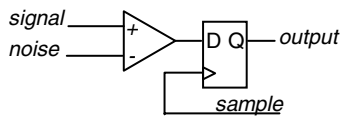

**Figure 6. Diagram of 1-bit digitizer**

The noise amplitude should be greater than or equal to the signal amplitude, and both signals should be zero mean (or have the same dc level). The autocorrelation at the output of the sampler, considering that the combined signal and noise is a normal stationary process with zero mean, is given by the arcsine law:

$$R_y(\tau) = \frac{2}{\pi} \arcsin\left(\frac{R_x(\tau)}{R_x(0)}\right) \quad (12)$$

Equation 12 allows one to state that the statistics of the input signal will be at the output of the sampler, affected by a gain factor and by the arcsine function, which is approximately linear for small values of the input argument.

As the Fourier transform of the autocorrelation is the power spectrum density, one is able to observe the spectral characteristics of the signal, but with an increased noise level because of the addition of the noise at the comparator input [3].

### 5.2 Evaluating power levels

In the following, a Matlab simulation illustrates the idea of measuring noise levels using a reference waveform. In a real application, noise and reference waveforms should be amplified in order to enable the use of a voltage comparator.

If one applies a constant-amplitude square wave signal to the digitizer of figure 6 as a reference signal, noise levels can be determined if a simple strategy is followed. Two noise levels were applied to the digitizer using the same square wave as reference. Signals are as shown in figure 7. One should notice that noise levels should be always greater than the reference levels.

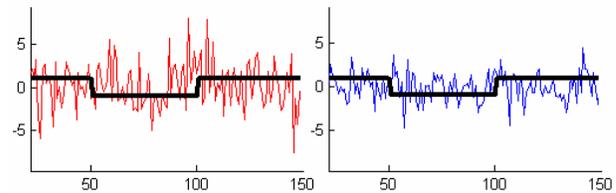

**Figure 7. Noise and reference waveforms for hot (left) and cold (right) noise temperatures**

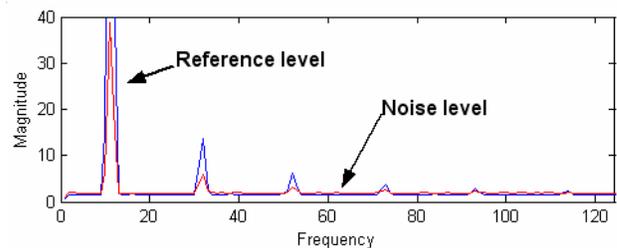

**Figure 8. Power spectrum density**

The power spectrum density evaluation of the bitstream at the output of the digitizer is shown in figure 8. One can notice that the noise levels remains similar, while amplitude levels of the reference square wave are larger.

As the reference level is constant, a simple normalization procedure can be used. One can evaluate the maximum amplitude of both spectra and apply a correction factor to one of the power spectral density plots.

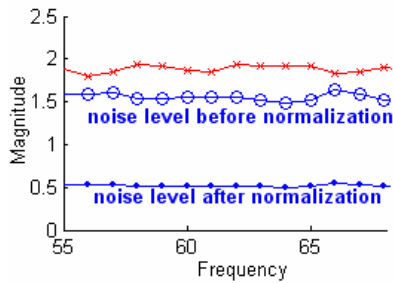

**Figure 9. Power spectrum density after normalization (*zoom* at 60 Hz)**

Figure 9 shows different noise levels, obtained before and after the normalization procedure. One can observe that the noise levels were very close before the normalization procedure. In order to make a numeric comparison, noise power ratio was evaluated using three different methods: ratio of mean square values (evaluated in time domain), ratio of PSD data and ratio of PSD data from the 1-bit digitizer. The values obtained are presented in table 2. For the 1-bit data, the reference waveform must be excluded from the power ratio evaluation (the reference is not part of the signal being measured). If this is accomplished, about 2.5% error in the power ratio was observed in the simulation, as presented by the last line in table 2.

**Table 2. Noise power ratio evaluation and derived parameters for $T_h$=10000K and $T_c$=1000K**

| Method | Noise power ratio | F | NF(dB) |
|---|---|---|---|
| Mean square ratio | 3.4866 | 10.03 | 10.01 |
| PSD ratio | 3.4766 | 10.08 | 10.03 |
| 1-bit PSD ratio **excluding** reference | 3.5620 | 9.66 | 9.85 |

### 5.3 Noise and reference levels

Simulations were carried out in order to evaluate the accuracy of noise power ratio estimates as a function of the amplitude of the reference waveform. Figure 10 shows the error in power ratio estimates for gaussian noise. The reference waveform amplitude level is related to the overall accuracy of the method: for very small amplitude references, a large error is expected because of noise levels disturbing the reference amplitude. Very large references may lead to non-linear distortion of the digitizer. Amplitudes in the range of 10% to 40% of the noise level should give reasonable results (figure 10).

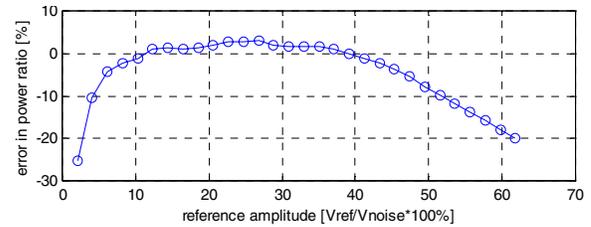

**Figure 10. Error in power ratio estimates versus reference amplitude**

### 5.4 Experimental results

In order to verify the approach a test setup was implemented in order to measure the noise figure of a non-inverting amplifier. The general setup is shown in figure 11. In order to change the value of the noise figure of the circuit, a different operational amplifier was used. As the equivalent noise voltages are provided by the data-sheets of the components, one is able to calculate the expected nominal value of the noise figure of the circuit, according to the opamp used [13].

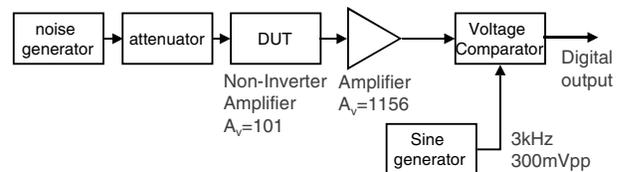

**Figure 11. Diagram of the experimental setup**

The opamps used were the OP27, OP07, TL081 and CA3140. The corresponding expected noise figures where in the range of 3.7dB to 16.2dB. In order to avoid interference pickup, the circuit was assembled in an aluminum case, being battery powered (figure 12).

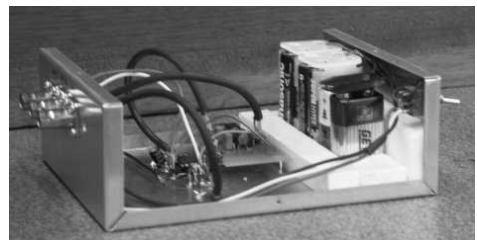

**Figure 12. Prototyped circuit**

External noise generator and sine wave generator (both HP33120A) were used. The reference waveform was at 3kHz, while the noise measurement bandwidth was at 1kHz, as indicated in figure 13. The output of the digitizer was acquired using a digital scope (HP54645D). Data was processed using Matlab. Total acquisition length was 1e6 samples and the FFT size was 1e4 samples. The results obtained after processing are presented in table 3, including the expected noise figure values from noise circuit analysis [13].

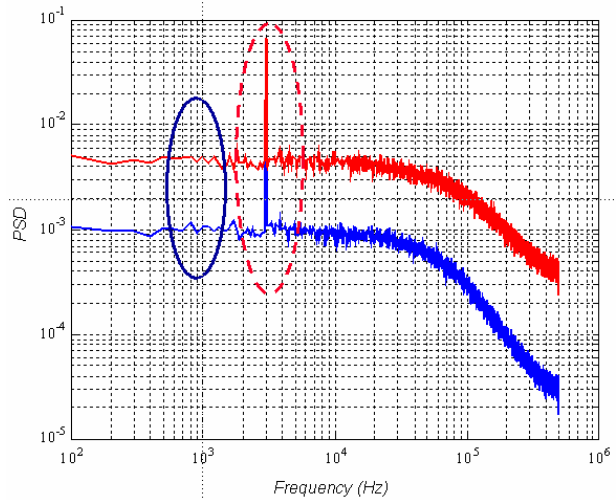

**Figure 13. PSD plot for noise levels after normalization**

**Table 3. Noise figure results for $T_0$=290K and $T_h$=2900K**

| Opamp | Expected | Measured |
|---|---|---|
| OP27 | 3.7 | 3.69 |
| OP07 | 6.5 | 4.841 |
| TL081 | 10.1 | 9.698 |
| CA3140 | 16.2 | 14.02 |

## 6. Analysis

The experimental results presented in section 5 have indicated the feasibility of implementing the approach for practical circuits. Noise figure measurements were carried out with a 2 dB maximum absolute error.

The proposed strategy makes use of a reference signal in order to perform the normalization process. Even a low-cost generator could be used, as the normalization process would track the main frequency component (disregarding harmonics, for example). This would enable the use of low quality reference waveforms, as the harmonics are not used in the normalization process. The amplitude of the main component, however, should be constant.

The need for analog signal conditioning, as a high gain amplifier, is related to the signal levels that should be measured. In general, an amplifier will be required for noise measurements [9]. This overhead could be minimized by observing that the noise figure of a cascade of stages is mainly the noise figure of the first stage [9].

## 7. Conclusions

In this paper a technique for the evaluation of noise figure in a BIST environment was presented. As the technique is based on DSP, it is able to benefit from resources already available in the SoC environment.

A simple voltage comparator and a reference signal were used as digitizers, thus discarding the need for an ADC. The technique also extends the capabilities of a simple BIST cell [3], allowing one to perform frequency and noise measurements.